\documentclass[twocolumn]{aastex7}

\begin{document}

\title{Thread Separation and Expansion Observed in Multi-Stranded Solar Coronal Loops}

\author[0000-0002-2189-9313]{David H. Brooks}
\affiliation{Computational Physics Inc., Springfield, VA 22151, USA}
\affiliation{University College London, Mullard Space Science Laboratory, Holmbury St. Mary, Dorking, Surrey, RH5 6NT, UK}
\affiliation{National Institutes of Natural Sciences, National Astronomical Observatory of Japan, 2-21-1 Osawa, Mitaka, Tokyo, 181-8588, Japan}
\email{dhbrooks.work@gmail.com}

\author[0000-0001-6102-6851]{Harry P. Warren}
\affiliation{Space Science Division, Naval Research Laboratory, Washington, DC 20375, USA}
\email{harry.warren@nrl.navy.mil}

\begin{abstract}

The theoretical expectation that coronal loops should expand with height contrasts with observations that typically show constant cross-sections. We investigate the idea that this discrepancy results from loops being composed of fine threads whose expansion occurs below the resolution limits of instruments like SDO/AIA. In this paper, we present two significant findings: (1) several extended loops exhibit measurable expansion, suggesting length as a critical factor in detection capability, and (2) high-resolution Solar Orbiter/EUI observations have captured expanding loops in active regions. For both AIA and EUI data, we observe cases where thread separation is directly visible as the loops evolve. These findings complement our previous work indicating AIA loops may consist of relatively few threads. Collectively, these observations provide compeling evidence supporting the multi-thread model and offer a potential resolution to the long-standing loop expansion problem in solar coronal physics. However, the high densities and narrow temperature distributions of observed coronal loops remain unresolved.

\end{abstract}


\section{Introduction} 

The magnetic field that rises above the photosphere is expected to expand to fill the coronal environment. Coronal loops that trace the magnetic field are 
therefore also expected to expand with height. EUV and X-ray observations of the corona, however, indicate that coronal loops have nearly uniform 
thickness along their lengths \citep{Klimchuk1992,Watko2000}. The explanation for this contradiction between theory and observations is a long standing puzzle.
Since the observed properties of coronal loops are directly related to their morphology and driven by the energy deposition process, the resolution of this
problem holds potential clues to the structure of coronal loops and the nature of the coronal heating mechanism.

Several novel explanations of these observations have been proposed including, for example, local twisting leading to constriction \citep{Klimchuk2000}, or expansion preferentially
along the line-of-sight leading to a selection effect \citep{Malanushenko2013}, but there is no consensus yet on the solution \citep{Klimchuk2020}. Most previous studies of 
coronal loop geometry, however, have focused on loops within active regions (ARs).
Here we conjecture that longer loops may exhibit more easily detectable expansion. Post-flare loop arcades expand over time such that loops formed
several hours after the flare peak time are more extended. Futhermore, it is well known that AR loops can interact and connect to neighboring 
ARs. Towards the peak of the solar cycle, the activity bands move to lower latitudes and interact \citep{McIntosh2015}. This forms very extended loops
that cross the equator. 
We therefore commenced a study to look for loop expansion by examining the cross-sectional variation along the lengths of long post-flare and trans-equatorial loops.

Another aspect of this problem is that until recently EUV imaging of the corona has generally only been possible with a spatial resolution of 1000\,km or more. In contrast, the solar
photosphere, the source of the coronal magnetic field, is structured on spatial scales of less than 100\,km.
Furthermore, although not direct measurements of coronal loops, observations of coronal rain - condensations that form in the catastrophic cooling phase of coronal loops - 
suggest spatial scales of a few hundred km \citep{Antolin2012} or even smaller \citep{Schmidt2025}. 
One possibility, therefore, is that loops are composed of fine threads whose expansion occurs below the spatial resolution of most current instrumentation \citep{DeForest2007}.

Many of the mechanisms proposed to heat coronal loops are expected to operate on spatial scales that are far below what can be achieved with these instruments. 
The fine threads composing coronal loops could therefore be very small \citep{Peter2013}. Conversely, it
remains an open question whether the fundamental size of observable coronal structures is close to current capabilities.
Several studies examining the properties of coronal loops have found that they mostly have narrow temperature distributions \citep{DelZanna2003,Aschwanden2005,Warren2008}, and that they
might only be composed of a few threads organized on spatial scales of a few hundred km \citep{Brooks2012}. This suggests that they are below the resolution limits of instruments
such as the Solar Dynamics Observatory \citep[SDO,][]{Pesnell2012} Atmospheric Imaging Assembly \citep[AIA,][]{Lemen2012}, but are close to being resolved. 
The highest spatial resolution EUV imaging to date has been achieved by the 
High-resolution Coronal Imager \citep[Hi-C,][]{Kobayashi2014,Rachmeler2019} on rocket flights in 2012 and 2018, and the 
Extreme Ultraviolet Imager \citep[EUI,][]{Rochus2020}
recently launched on Solar Orbiter \citep{Muller2020}. These instruments have a spatial resolution 3--5 times better than AIA. 
Various studies of Hi-C flight data have also suggested that coronal loops are close to being resolved \citep{Aschwanden2017,Williams2020a,Williams2020b}.

If loops are composed of fine threads whose expansion occurs on spatial scales that are below the resolution limits of SDO/AIA, it might therefore now be possible to 
resolve some of the individual threads and potentially detect the expansion
in AR loops with higher spatial resolution observations from EUI. 
In principle we expect quite significant expansion. The loop width should scale inversely with the coronal magnetic field strength, which itself is expected to fall
off approximately with the inverse of the radius cubed \citep{Klimchuk2000}. Detailed analysis of potential field extrapolations of both quiescent and flare-active ARs 
suggest that area expansion factors increase substantially with height. See, for example, the reproduction of results from \cite{Dudik2014} and \cite{Ugarte2007} by 
\cite{Reep2022}, which shows area expansion factors of 100 or more at heights $>$ 100\,Mm. 

The geometry of coronal emission is one important component of the larger problem of reconciling the properties of the solar upper atmosphere with physical models. The high densities and extended cooling times of the million-degree corona have proven impossible to reproduce with traditional stranded loop models \citep[see][]{Klimchuk2006}. This suggests that understanding the complex substructure in observed coronal loops is essential to resolving this mystery. \cite{Malanushenko2022} have
made the more radical suggestion that what appear to be coronal loops are actually complex, sheet-like structures, and ultimately this may prove
to be the case. Here we implicitly assume that coronal emission is organized into loop structures. 

This paper is organized in the following way.
We initially describe 
our results on long post-flare and trans-equatorial
loops observed by AIA. We show examples of long loops where expansion is clearly observed. 
We then discuss the new higher spatial resolution observations of shorter AR loops from EUI, and show examples of expansion in AR loops that is detected
by the new high resolution observations. Furthermore, for both AIA and EUI, we present examples where thread separation is clearly visible as the loops
evolve. Finally, we discuss a simple geometrical model of the observed loop expansion that explores the effect of the instrument point spread function (PSF)
on its detectability with an instrument such as AIA. 

\section{Data analysis method} 

For this investigation, we used data from SDO/AIA
and Solar Orbiter/EUI. AIA observes in several wavelength bands with a spatial resolution 
of 0.6$''$/pixel. EUI has two High Resolution Imagers (HRI) that observe with a spatial resolution of 0.5$''\times d_{\sun}$/pixel, where $d_{\sun}$ is the 
distance between the Sun and Solar Orbiter at the time of the encounter. 
For comparison between the two instruments, we focus on observations with the AIA 171\,\AA\, and EUI 174\,\AA\, filters. 
These AIA and EUI filters are dominated by emission from strong spectral lines of \ion{Fe}{9} and \ion{Fe}{10} forming near 0.8--1.1\,MK.
Exposure times for images from both instruments are 2\,s.

We use data calibrated by the instrument teams. The AIA data were downloaded from the Lockheed Martin cutout service. The data are designated level 1.5 and have been corrected for
dark current and flat fields, and cleaned of cosmic rays and bad pixels.
We obtained the EUI data from the Solar Orbiter data archive at ESA. These data are designated level 2 and have been corrected for instrumental issues with the best
available procedures at the time of the data release. We use data release 6.0 in this work \citep{Kraaikamp2023}.

\begin{figure*}[ht!]
    \includegraphics[viewport = 100 150 542 552,clip=true,width=0.5\textwidth]{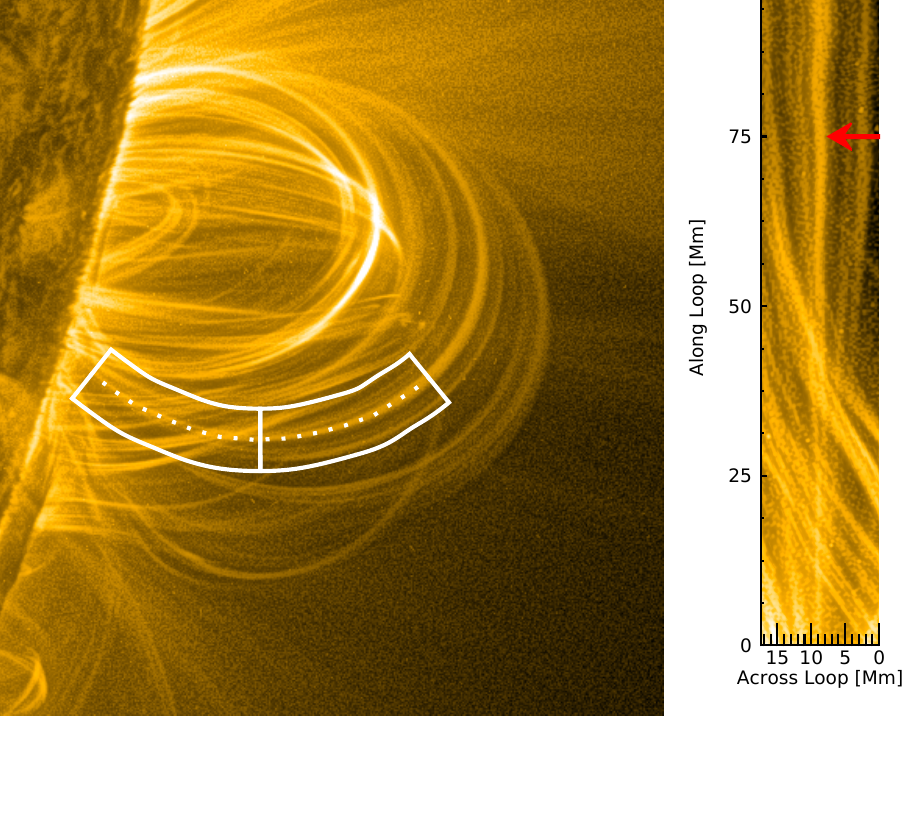}
    \includegraphics[viewport = 120 180 512 552,clip=true,width=0.5\textwidth]{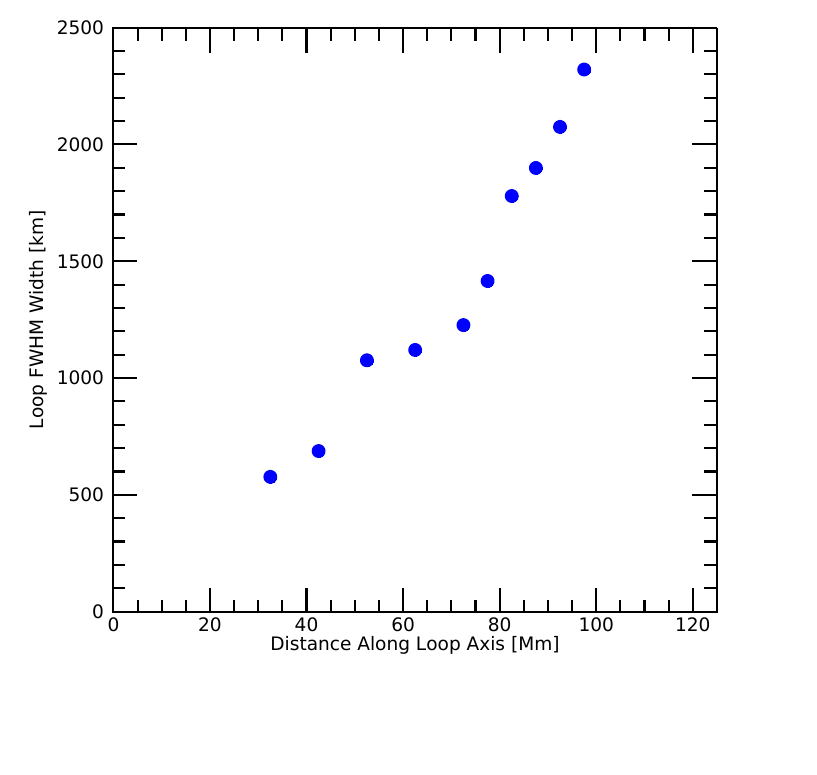}
  \centerline{
    \includegraphics[viewport= 20 330 612 432,width=1.0\textwidth]{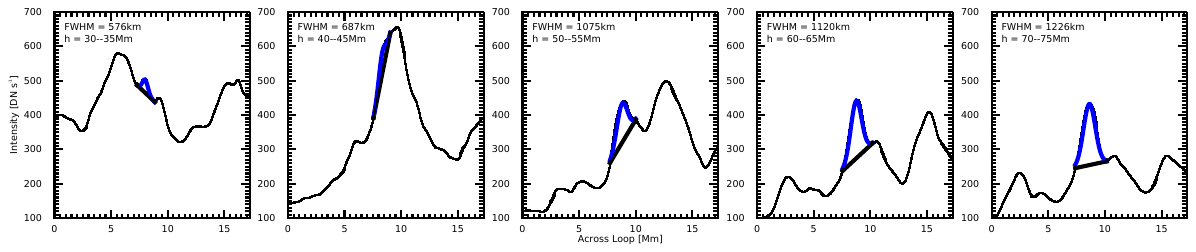}}
  \centerline{
    \includegraphics[viewport= 20 330 612 432,width=1.0\textwidth]{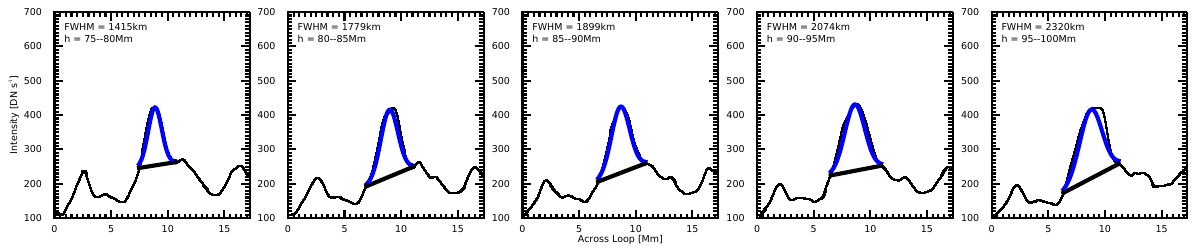}}
\caption{Post-flare loop observed in the AIA 171\,\AA\, filter on 2012 July 19, at 12:03\,UT. 
The solid box in the upper left panel outlines
the segment of the loop that was traced. The dotted line follows the loop. The upper middle panel shows 
the straightened loop segment on a Mm spatial scale. The red arrow highlights the straightened loop. The wave-like pattern towards the end of
the loop is a result of imperfect straightening over such large curved distances. 
These and subsequent images in other figures have been sharpened using the MGN procedures of \cite{Morgan2014}.
We used these images to aid loop identification, but all of the analysis was performed on the original images. 
The upper right panel shows measurements of the loop width (FWHM) as a function of distance along the post-flare loop. The data points correspond
to the cross-field intensity profiles also shown as a function of distance in the lower two rows. In these plots, the
raw data are shown by the solid black line. The gaussian fits to the cross-loop profiles are shown in blue. The
solid straight line shows the fit between the selected background positions.
The distance of the measurement from the start of the trace and the width of the loop at that position are shown in the legend.
\label{fig1}}
\end{figure*}

\begin{figure*}[ht!]
\includegraphics[viewport = 100 150 542 552,clip=true,width=0.5\textwidth]{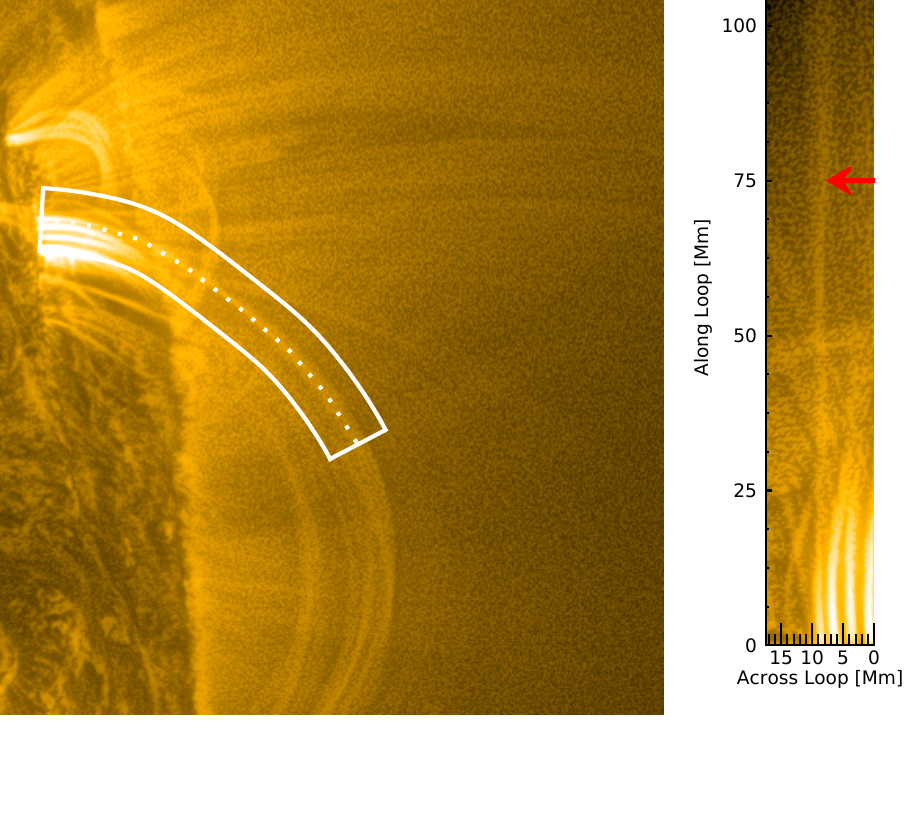}
\includegraphics[viewport = 120 180 512 552,clip=true,width=0.5\textwidth]{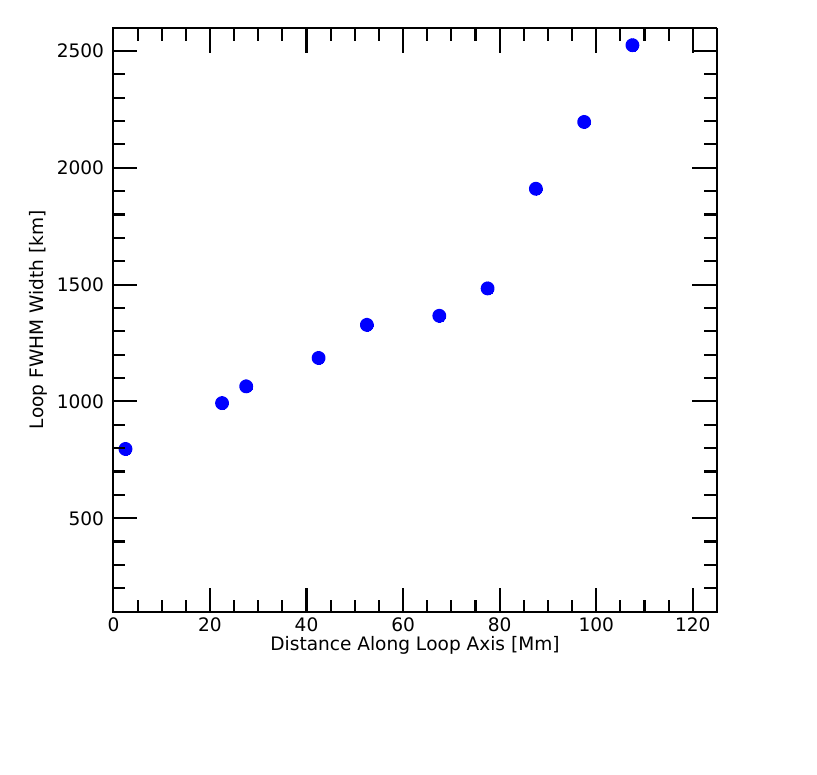}
  \centerline{%
    \includegraphics[viewport= 20 330 612 432,width=1.0\textwidth]{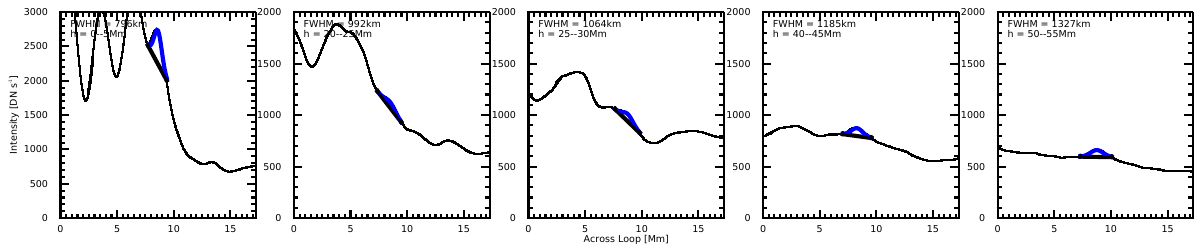}} %
  \centerline{%
    \includegraphics[viewport= 20 330 612 432,width=1.0\textwidth]{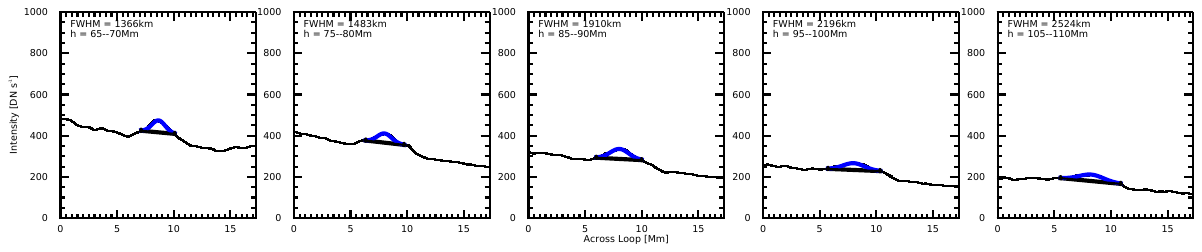}} %
\caption{Same as Figure \ref{fig1} but showing a trans-equatorial loop observed in the AIA 171\,\AA\, filter on 2025 April 1, at 08:00\,UT.
\label{fig2}}
\end{figure*}
\begin{figure*}[ht!]
\includegraphics[viewport = 100 150 542 552,clip=true,width=0.5\textwidth]{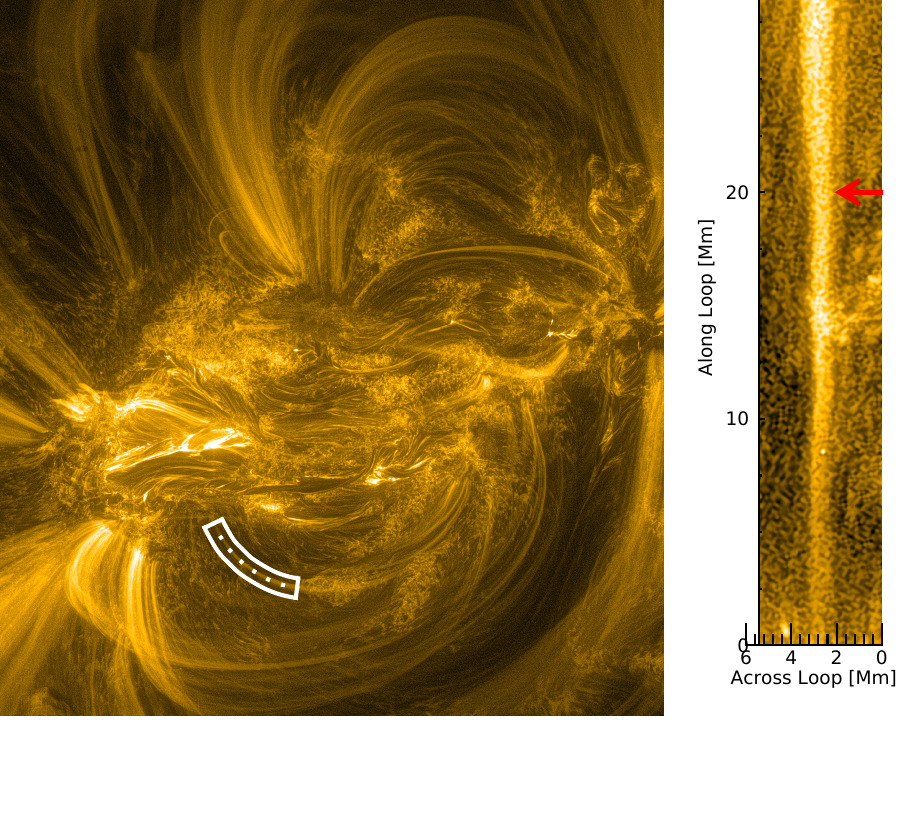}
\includegraphics[viewport = 120 180 512 552,clip=true,width=0.5\textwidth]{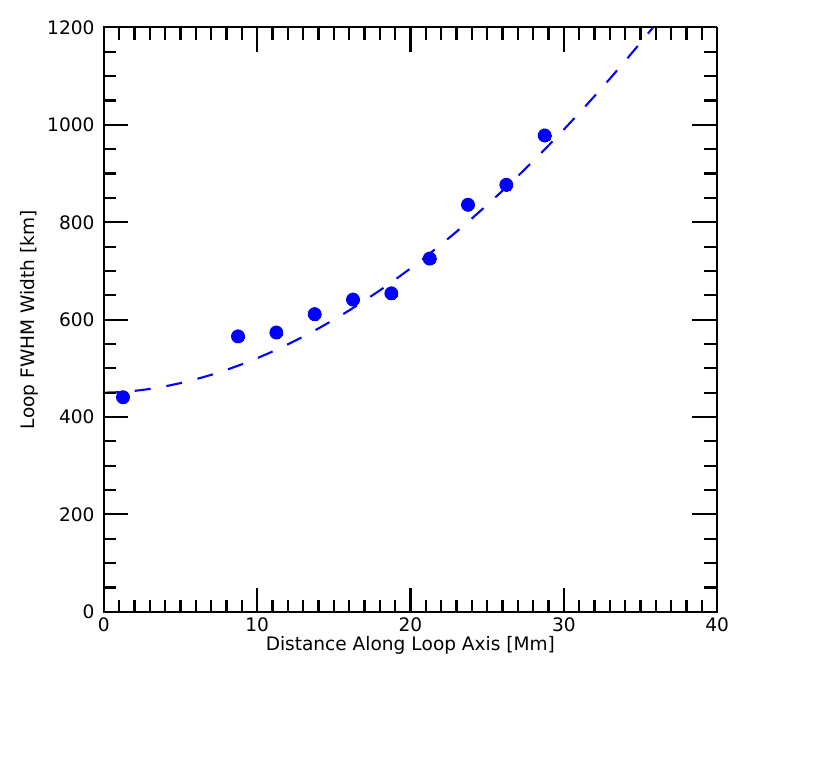}
  \centerline{%
    \includegraphics[viewport= 20 330 612 432,width=1.0\textwidth]{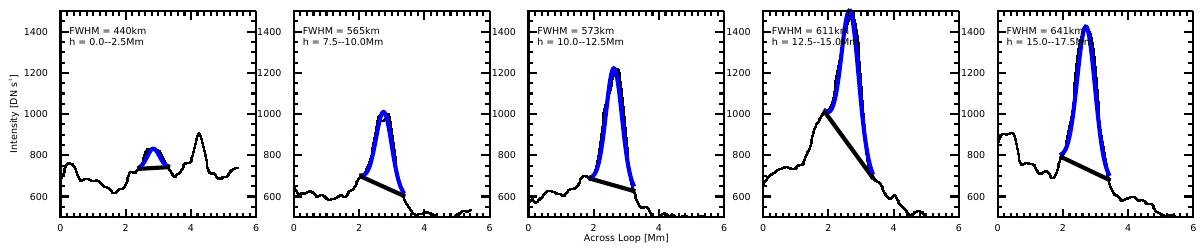}} %
  \centerline{%
    \includegraphics[viewport= 20 330 612 432,width=1.0\textwidth]{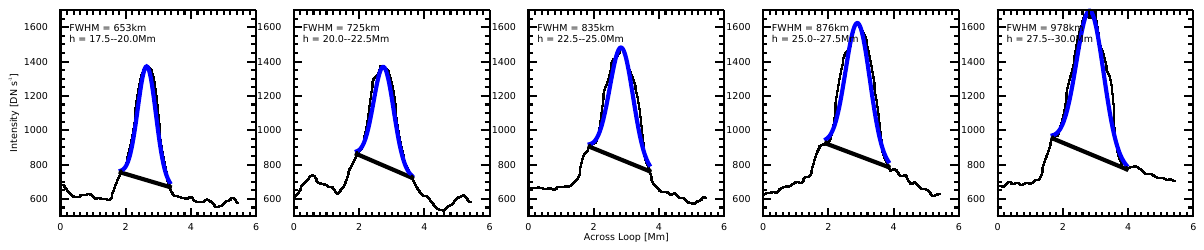}} %
\caption{Same as Figure \ref{fig1} but showing an active region loop observed in the EUI 174\,\AA\, filter on 2024 March 23, at 22:55\,UT.
The spatial resolution of the EUI observations on this date was 0.188$''$/pixel (see text).
Note that for this shorter loop the measurements are made at intervals of 2.5\,Mm. The dashed line in the top right panel is a polynomial
fit to the loop widths measured in the lower panels. It is used in Section \ref{model} for an investigtion of the effect of the instrument
PSF on the detection of the observed expansion.
\label{fig3}}
\end{figure*}
\begin{figure*}[ht!]
\includegraphics[viewport = 100 150 542 552,clip=true,width=0.5\textwidth]{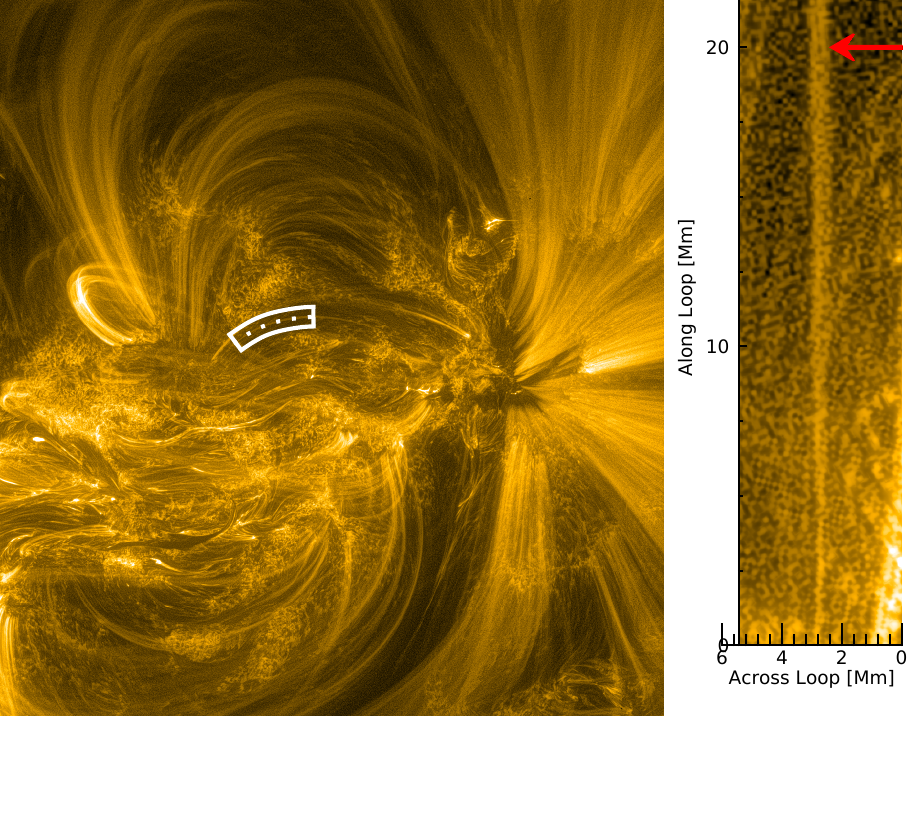}
\includegraphics[viewport = 120 180 512 552,clip=true,width=0.5\textwidth]{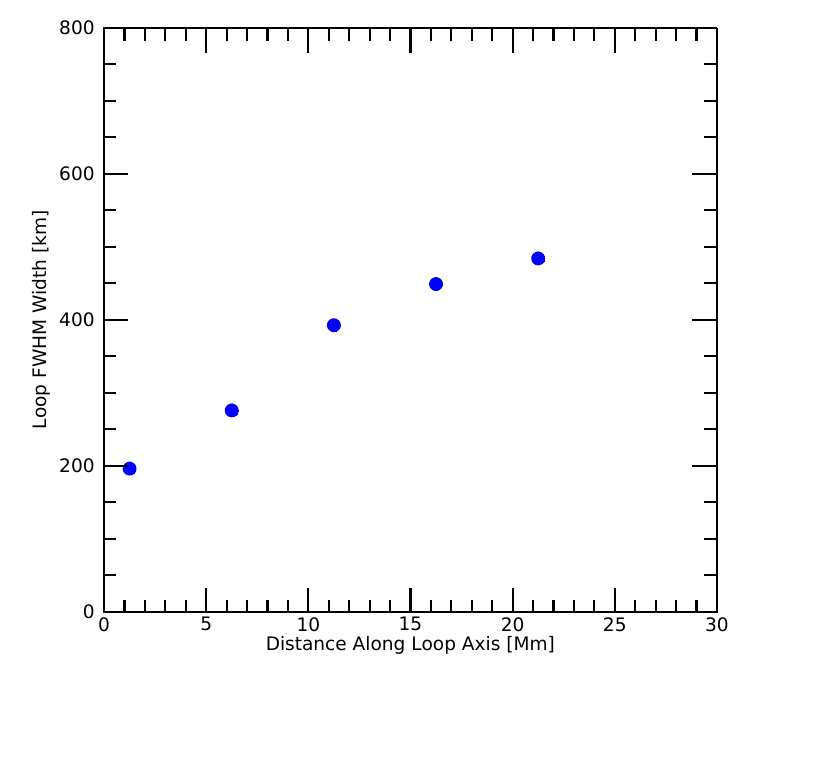}
  \centerline{%
    \includegraphics[viewport= 20 330 612 432,width=1.0\textwidth]{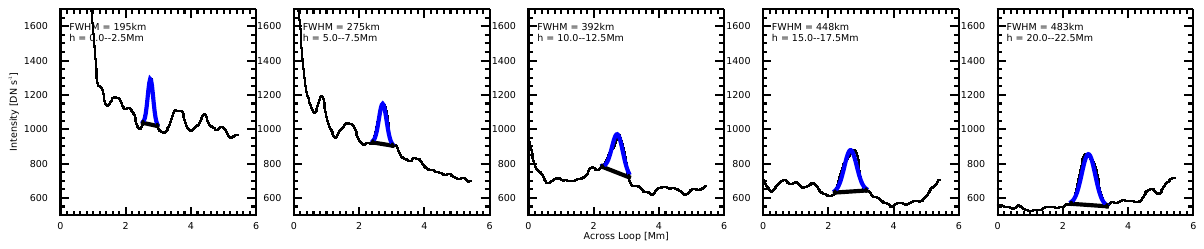}} %
\caption{Same as Figure \ref{fig1} but showing another active region loop observed in the EUI 174\,\AA\, filter on 2024 March 23, at 23:15\,UT.
As in Figure \ref{fig3}, the measurements for this shorter loop are made at intervals of 2.5\,Mm.
\label{fig4}}
\end{figure*}

To aid in loop identification we applied the multi-scale Gaussian normalization (MGN) procedures of \cite{Morgan2014} to the AIA and EUI images.
Once we identify a candidate loop in the images, we extract a straightened image of the loop along its axis. We then measure the loop width within 5\,Mm segments
(2.5\,Mm for EUI)
at different positions along the straightened loop. Note that all of the quantitative analysis was performed on the raw images. 
This is achieved by extracting the cross-loop intensity profiles perpendicular to the loop, 
and averaging these profiles along the loop segment. We then identify the loop in the intensity profile, and subtract a co-spatial background 
by fitting a first-order polynomial between two visually selected background pixels. The width of the Gaussian component of the polynomial is then converted
to a full-width half-maximum (FWHM) value. The procedure is then repeated for each segment along the straightened loop.
The method is a modification of well established tehniques that have been used extensively in previous studies of properties of active region loops
such as temperatures
\citep{Aschwanden2008a,Warren2008}, spatial scales \citep{Brooks2012}, and coronal magnetic field strengths \citep{Brooks2021}.

\begin{figure*}[ht!]
  \begin{interactive}{animation}{fig5_animation.mp4}
  \centerline{%
    \includegraphics[viewport= 20 270 612 502,width=1.0\textwidth]{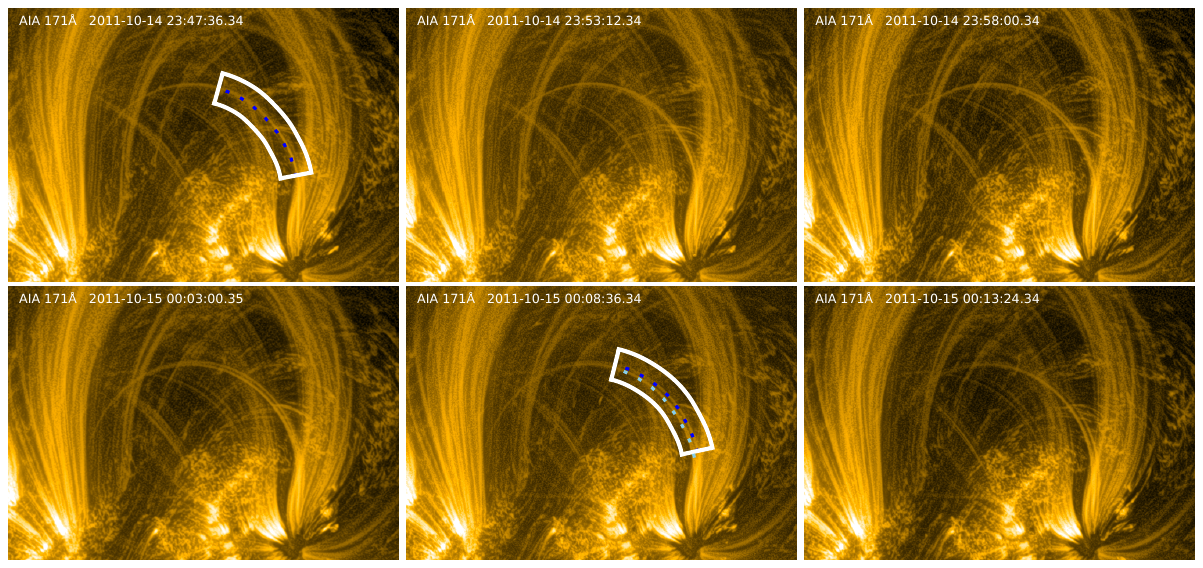}} %
  \centerline{%
    \includegraphics[viewport= 20 330 612 462,width=1.0\textwidth]{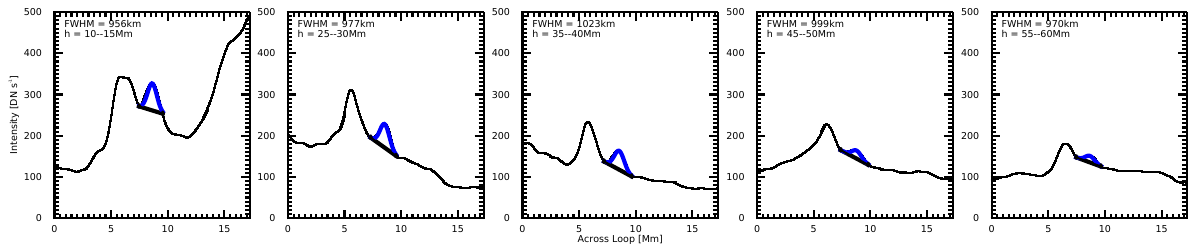}} %
  \centerline{%
    \includegraphics[viewport= 20 330 612 462,width=1.0\textwidth]{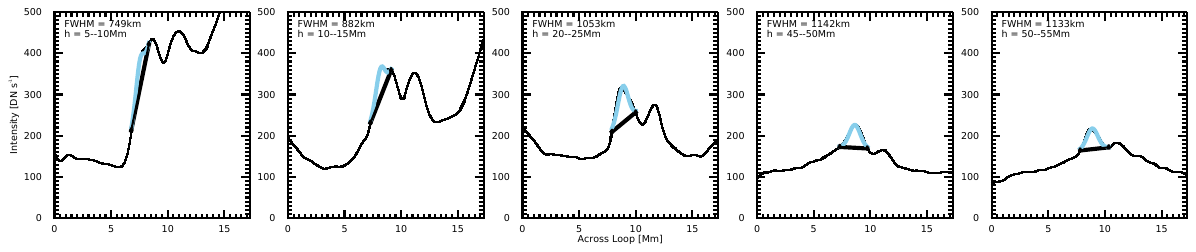}} %
  \end{interactive}
\caption{ AIA 171\,\AA\, images showing an example of a loop separating into two distinct threads.
The solid box in the upper left panel shows the area encompassing the segment of the single loop that was traced.
The blue dotted line follows the loop.
The solid box in the middle panel of the second row outlines the area encompassing the 
segments of the two separated threads that were traced. The dotted lines follow the threads. They are color coded blue and sky blue
for cross-referencing with the intensity profiles in the lower two rows. 
The cross-field intensity profiles as a function of distance along the separating threads are shown in the lower two rows.
The raw data are shown by the solid black line. The gaussian fits to the cross-thread profiles are color coded blue and sky blue
for cross-referencing with the traces in the images. The
solid straight line shows the fit between the selected background positions.
The distance of the measurement from the start of the trace and the width (FWHM) of the thread at that position are shown in the legend.
The animation associated with this figure runs for 22\,s and covers the period from 2011 October 14 21:00\,UT to
2011 October 15 02:00\,UT. It shows a single panel from the static figure with the same FOV. It is annotated with an arrow to indicate
the loop that separates into two threads.
\label{fig5}}
\end{figure*}

\section{Observations of Loop Expansion}
We illustrate the application of our analysis procedure to a number of loops in Figures \ref{fig1}--\ref{fig4}.

\subsection{Long post-flare loops}
An M7.7 flare (SOL2012-07-19) occurred at the west limb in AR 11520 on 2012 July 19. The GOES soft X-rays peaked at 05:58\,UT, but the event 
lasted several hours. Post-flare loops had cooled enough to be visible in the AIA 171\,\AA\, filter as early as 05:20\,UT.
Loops in the post-flare arcade formed on relatively short spatial scales (tens of Mm) initially, but expanded over the next
few hours. Figure \ref{fig1} shows an example of one of these long loops at 12:03\,UT. We traced a significant portion of this loop for analysis.
The straightened loop segment shows that the loop half length exceeds 100\,Mm. It is already clear from the straightened image that the loop
expands significantly.

We measured the loop width in 5\,Mm segments along the straightened loop. Figure \ref{fig1} also shows examples of the cross-loop
intensity profile at different positions/heights, $h$, along the loop. We also show the measured width at each position. 
The width is 576\,km at 30-35\,Mm, and increases to 2320\,km at 95-100\,Mm: an expansion of a factor of 4.

\begin{figure*}[ht!]
  \begin{interactive}{animation}{fig6_animation.mp4}
  \centerline{%
    \includegraphics[viewport= 20 270 612 502,width=1.0\textwidth]{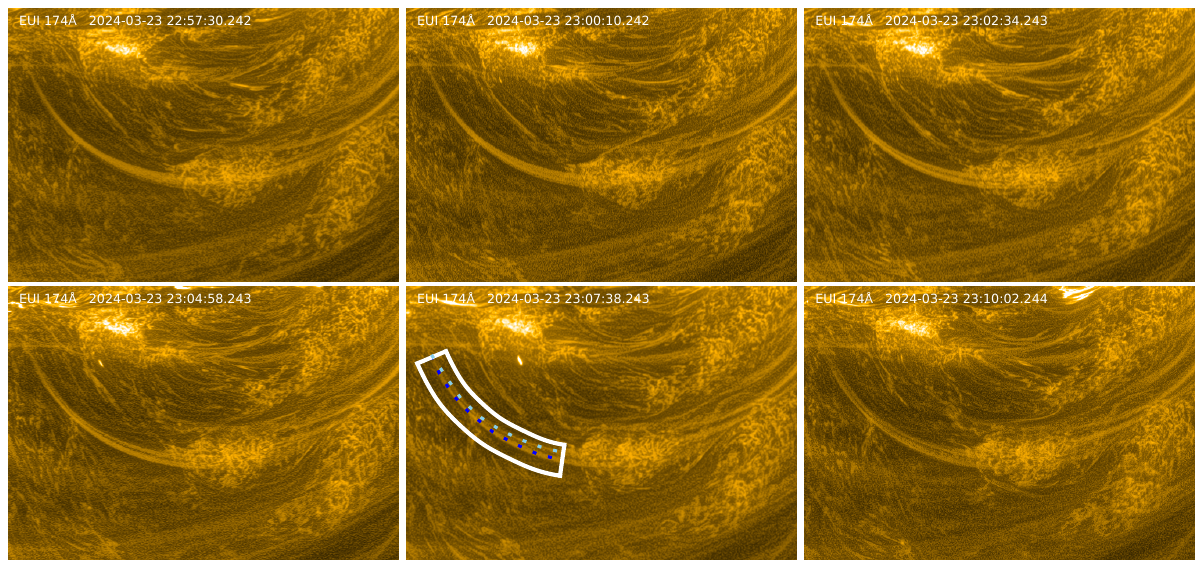}} %
  \centerline{%
    \includegraphics[viewport= 20 330 612 462,width=1.0\textwidth]{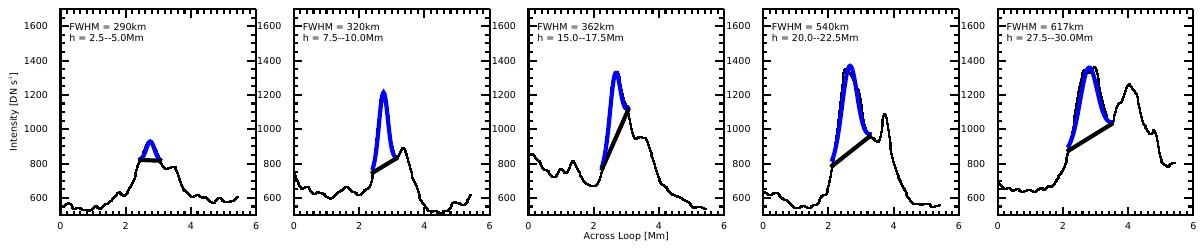}} %
  \centerline{%
    \includegraphics[viewport= 20 330 612 462,width=1.0\textwidth]{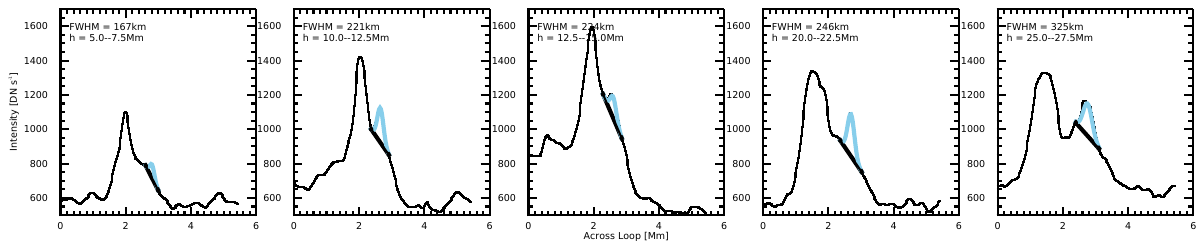}} %
  \end{interactive}
\caption{ Same as Figure \ref{fig5}, but showing EUI 174\,\AA\, images of the loop in Figure \ref{fig3} separating into two distinct threads.
The animation associated with this figure runs for 19\,s and covers the period from 2024 March 23 22:30\,UT to
2024 March 24 00:30\,UT. It shows a single panel from the static figure with the same FOV. It is annotated with arrows to indicate
the loops from Figure \ref{fig3} and Figure \ref{fig7} that separate into two threads.
\label{fig6}}
\end{figure*}

Note that the values of $h$ are relative to the base of the loop trace rather than the actual loop footpoint.
We stress that these measurements are difficult. Line of sight confusion between different structures makes it hard to trace 
individual loops cleanly. First, faint emission at the loop top makes it more difficult to separate the loop from the background 
emission. Second, several loops are often rooted close to each other at their base and the AIA spatial resolution 
limitations make it hard to visually separate them. This is the case in our example in Figure \ref{fig1}.
We are showing what we consider to be a fairly clear example of loop expansion, but even in this case, there is uncertainty in
the true expansion factor because of line-of-sight 
confusion by another loop at lower heights. It appears that we have two loops rooted close to each other at the footpoints,
but their different trajectories allow us to see them separated at around 30-35\,Mm (straightened loop image in Figure \ref{fig1}).
If we extend our measurements down to 0--5\,Mm, we obtain a width of 719\,km, but we are likely measuring the widths of two loops. This implies that the
measurement could be too large, and thus the lower inferred expansion factor (3.2) could be too small. Conversely, we could end up tracing our 
loop too low, in this case implying that the measured width at 0--5\,Mm is from the second loop. If the main loop continues to contract, our
inferred expansion factor (4) could also be too low. 

Despite this discussion of limitations, it is clear that this post-flare loop shows significant expansion.

\begin{figure*}[ht!]
  \centerline{%
    \includegraphics[viewport= 20 270 612 502,width=1.0\textwidth]{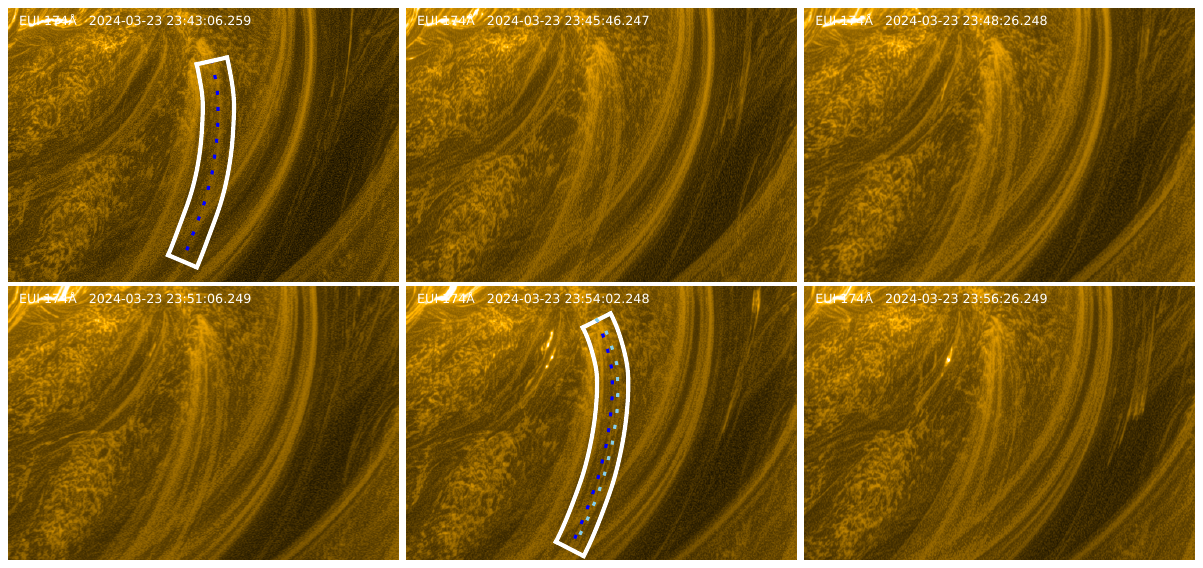}} %
  \centerline{%
    \includegraphics[viewport= 20 330 612 462,width=1.0\textwidth]{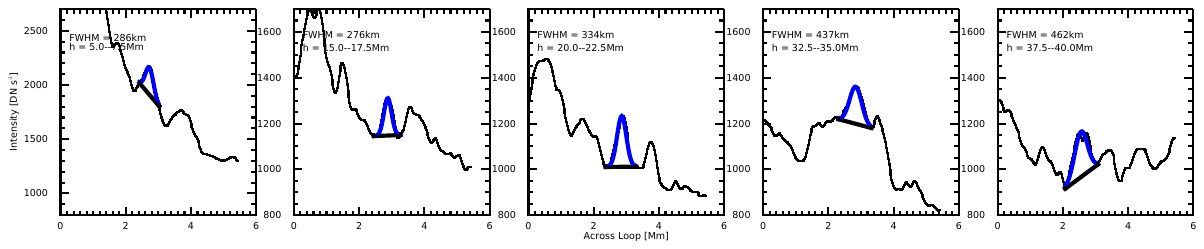}} %
  \centerline{%
    \includegraphics[viewport= 20 330 612 462,width=1.0\textwidth]{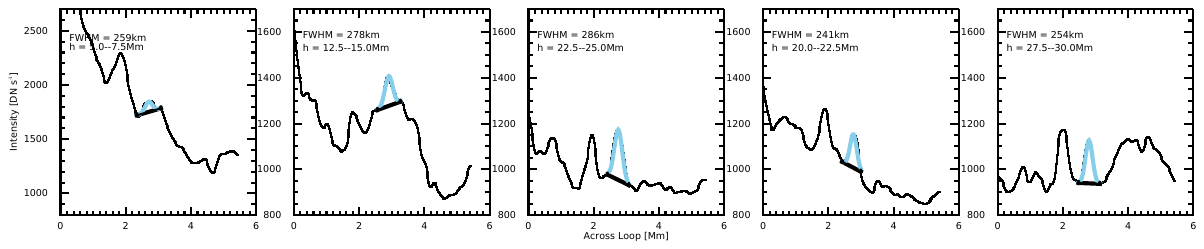}} %
\caption{ Same as Figure \ref{fig5}, but showing EUI 174\,\AA\, images of a different example of a loop separating into two distinct threads.
This example also appears in the animation associated with Figure \ref{fig6}.
\label{fig7}}
\end{figure*}

\subsection{Long trans-equatorial loops}
We are now close to the peak of solar cycle 25 and trans-equatorial coronal loops are seen fairly regularly. On 2025 April 1 extended
loops were observed on the west limb connecting AR 14039 in the northern hemisphere with a complex of several ARs in the southern hemisphere.
Figure \ref{fig2} shows an example of one of these long loops at 08:00\,UT. As with the post-flare loop, we traced a segment along the loop
for analysis. In this case, it was possible to trace the loop to the footpoint, but the reduced emission and line-of-sight confusion near the top 
made it difficult to cleanly trace the loop to the apex. The straightened loop segment shows 
that the loop half length again exceeds 100\,Mm. 

We measured the loop width in 5\,Mm segments along the straightened loop. Figure \ref{fig2} also shows examples of the cross-loop
intensity profile at different positions, $h$, along the loop. We also show the measured width at each position. 
The width is 796\,km at 0-5\,Mm, and increases to 2524\,km at 105-110\,Mm: an expansion of a factor of 3.2.

To give an idea of how this trans-equatorial loop and the earlier post-flare loop compare to previous studies, we note that only 1 of the 
30 AR loops in the survey by \cite{Aschwanden2008a} had a half length exceeding 100\,Mm. The longest length in the study of 23 loops by \cite{Klimchuk2020}
is 62.1\,Mm, and they state that their measured loop segments constitute $``$the majority of the total loop length, but not always a large majority$"$.
This suggests that they cover at least the loop half length and that most, if not all, have a full length less than 100\,Mm. 
Since we did not measure the complete footpoint-apex length of our two loops,
it is likely that they are longer than any of the loops in the \cite{Aschwanden2008a} and \cite{Klimchuk2020} studies. Some of the earlier surveys that
did not find any expansion in the longer loops in their samples probably did include a few loops as long as the ones investigated here. \cite{Klimchuk2000},
for example, looked at 43 loops and 14\% of the sample had half lengths exceeding 100\,Mm, with perhaps 2 being comparable to the loops we examined. A similar
sample was analyzed by \cite{Watko2000}.

\subsection{AR loops detected by high resolution observations}
We have shown clear examples of loop expansion in extended loops observed by AIA. We now examine whether the higher spatial resolution of Solar Orbiter/EUI 
enables us to detect expansion in shorter AR loops.

On 2024 March 23 Solar Orbiter was close to the Sun-Earth line at a distance of $d_{\sun}$ = 0.381\,AU. EUI thus observed AR 13615 with a spatial resolution
of 0.188$''$/pixel, a factor of 3.2 better than AIA. 
Figure \ref{fig3} shows an example image, taken at 22:55\,UT, of an AR loop that appears to exhibit expansion. We traced a segment along this loop
for analysis. This is a shorter loop than the post-flare and trans-equatorial loops we analyzed. The straightened loop segment is about 30\,Mm in length, though 
the loop is not followed to the apex. The half length is probably closer to 50\,Mm.

Since the loop is shorter, we measured the loop width in 2.5\,Mm segments along the straightened loop. Figure \ref{fig3} also shows examples of the cross-loop
intensity profile at different positions, $h$, along the loop. We also show the measured width at each position. 
The width is 440\,km at 0-2.5\,Mm, and increases to 978\,km at 27.5-30\,Mm: an expansion of a factor of 2.2.

Figure \ref{fig4} shows an image, taken at 23:15\,UT, of another example AR loop that appears to exhibit expansion. We also traced a segment along this loop
for analysis. The straightened loop segment in this case is about 25\,Mm in length. Our trace likely is close to the half length in this case.
Since this loop is even shorter than the one in Figure \ref{fig3}, we could not measure its width in as many 2.5\,Mm segments, so Figure \ref{fig4} shows examples of the cross-loop
intensity profile at only five different positions, $h$, along the loop. 
The width is 195\,km at 0-2.5\,Mm, and increases to 483\,km at 20-22.5\,Mm: an expansion of a factor of 2.5.

\section{Observations of loop thread separation}
Our previous work \citep{Warren2008,Brooks2012} suggested that loops observed by AIA are composed of finer magnetic threads, but that the number of threads
may be relatively small. The detection of expansion in AR loops by EUI supports the suggestion that at least some loops are now resolved. Another observation
that supports this view is that we observe loops where separation into distinct individual threads occurs while the loops evolve. 

We show an example of a coronal loop observed by AIA that clearly separates into two distinct threads in Figure \ref{fig5}. The figure shows the evolution of
the loop and separation of the threads in a series of example images. 
The loop is visible as a single structure in the first image at 23:47:36\,UT, but starts to separate into two threads 
from near its apex in the third image taken at 23:58:00\,UT. The two threads are visibly separated from the apex to the footpoints in the subsequent images. 
The initial single structure shows only limited expansion. Following the same analysis
as before (not shown here for brevity), we measured a width of 963\,km at 5-10\,Mm, that increases to 1361\,km at 60-65\,Mm: an expansion of 1.4

We measured the widths in 5\,Mm segments along the straightened separated threads. Figure \ref{fig5} also shows examples of the cross-loop
intensity profile at different positions, $h$, along the two threads. We also show the measured width at each position. 
The width is 956\,km at 10-15\,Mm, and remains broadly constant with a value of 970\,km at 55-60\,Mm\, for the thread in the upper panel. 
The width is 749\,km at 5-10\,Mm, and increases to 1133\,km at 50-55\,Mm\, for the thread in the lower panel: an expansion of a factor of 1.5.
Note that the lower thread, sky blue in the middle panel of the second row in Figure \ref{fig5}, appears to separate into two near its footpoint. We traced the lowest
thread at the footpoint to avoid an artificially large measurement.
It is interesting that the expansion factor of the thread in the lower panel is quite similar to the value for the original unseparated loop. 
This loop was previously studied by \cite{Brooks2012}.

The expanding AR loop we highlighted in Figure \ref{fig3} also shows clear thread separation. Figure \ref{fig6} shows the evolution of
the loop and separation of the threads in a series of example images. 
The loop is visible as a single expanding structure in the first image at 22:57:30\,UT, but starts to separate into two threads 
from near its apex in the second image at 23:00:10\,UT. The two threads are completely separated along their full lengths by the time of the 23:07:38\,UT\, image.

We measured the widths in 2.5\,Mm segments along the straightened separated threads. Figure \ref{fig6} also shows examples of the cross-loop
intensity profile at different positions, $h$, along the two threads. We also show the measured width at each position. 
The width is 290\,km at 2.5-5\,Mm, and increases to 617\,km at 27.5-30\,Mm\, for the thread in the upper panel: an expansion of a factor of 2.1.
The width is 167\,km at 5-7.5\,Mm, and increases to 325\,km at 25-27.5\,Mm\, for the thread in the lower panel: an expansion of a factor of 1.9.

Although the 
measurements used for this comparison were not taken at exactly the same positions along each of the structures, it is interesting that the 
thread expansion factors of 2.1 and 1.9 are close to the result for the original unseparated loop (2.2). Futhermore, the combined width of the two threads 
(457\,km at 2.5-7.5\,Mm and 942\,km at 25-30\,Mm) is also
close to the result for the original loop (440\,km at 0-2.5\,Mm and 978\,km at 27.5-30\,Mm). 
This is very much consistent with
the idea that the original loop really is composed of only a few threads.

We show another example in a series of images in Figure \ref{fig7}.
The loop is visible as a single structure in the first image at 23:43:06\,UT, but separates into two threads along its length that are visible in the
image at 23:54:02\,UT. This example is similar to the AIA one. The initial single structure shows limited expansion. Following the same analysis
as before (not shown here for brevity), we measured a width of 350\,km at 0-2.5\,Mm, that increases to 558\,km at 30-32.5\,Mm: an expansion of a factor of
1.6. 

We also show the cross-loop intensity profile at different positions, $h$, along the two separated threads in Figure \ref{fig7}. 
The width is 286\,km at 5-7.5\,Mm, and increases to 462\,km at 37.5-40\,Mm\, for the thread in the upper panel: an expansion of a factor of 1.6.
Like the AIA example, the other thread does not show expansion.
The width, for the thread in the lower panel, is 259\,km at 5-7.5\,Mm, and remains broadly constant with a value of 254\,km at 27.5-30\,Mm. 

\begin{figure*}[ht!]
  \centerline{%
    \includegraphics[width=1.0\textwidth]{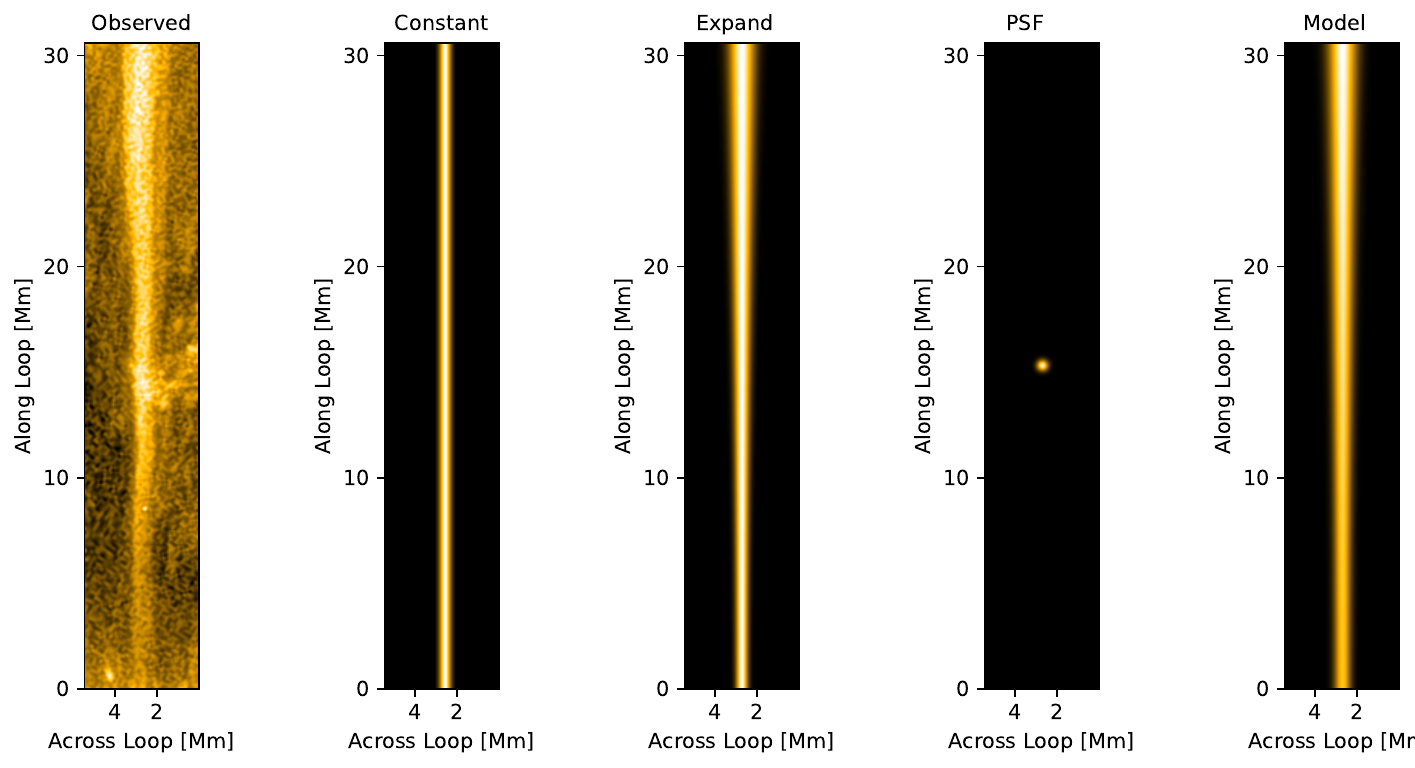}} %
\caption{ A simple model of the impact of the PSF on detection of the expansion of the EUI loop shown in Figure \ref{fig3}.
Left to right: a reproduction of the straightened image of the EUI loop from Figure \ref{fig3}. A simple constant cross-section model of this loop
built by taking the measured FWHM at the loop base, and assuming a cross-loop Gaussian intensity profile with a constant width along the loop. 
A simple model of the loop expansion, built by taking the FWHM measured at the loop base, and assuming a cross-loop Gaussian intensity profile with an
increasing width along the loop. The width increase is from a polynomial fit to the measured FWHMs along the loop in Figure \ref{fig3} (top right panel). 
A 2-D Gaussian model of the AIA PSF with a FWHM of 434\,km. The model of the EUI expanding loop after convolution with the AIA PSF.
\label{fig8}}
\end{figure*}

Animations of the regions of the ARs covering these thread separation examples are available
in the electronic version of the manuscript. 

\section{Modeling of loop expansion}
\label{model}
The observations we have reported suggest a coherent picture where loops are composed of fine threads on spatial scales that are
just below the resolution of AIA. Their expansion is therefore difficult to detect, but can be seen in long post-flare and trans-equatorial
loops. The expansion can also sometimes be detected in the higher spatial resolution of EUI observations. If this picture is correct, 
a question arises as
to how narrow the loops need to be to avoid detection of the loop expansion when convolved with the instrument PSF? Here we 
perform some comparisons with simple models to try to address this question and check for consistency with the observations.

Figure \ref{fig8} shows a simple model of the effect of convolving the expanding EUI loop studied in Figure \ref{fig3} with a 2-D Gaussian
model of the AIA PSF. The figure shows a straightened image of the EUI loop segment on a Mm spatial scale, and then a constant cross-section
model of the loop assuming a Gaussian cross-loop intensity profile with a FWHM comparable to that measured at the loop base in the top right
panel of Figure \ref{fig3}. The third panel shows a simple model of the loop expansion, again assuming a Gaussian cross-loop intensity profile,
but this time increasing the width along the loop using a polynomial fit to the measured FWHMs along the loop shown as the dashed line in the top right panel of
Figure \ref{fig3}. The polynomial fit takes the form $w(d) = w_0 + d^\alpha$, where $w(d)$ is the FWHM as a function of distance, $d$, along the loop, $w_0$
is the FWHM at the loop base, and $\alpha$ is 1.85. 

We remind the reader that the measured expansion of this EUI loop is a factor of 2.2, and this is confirmed by applying the same measurement 
technique as we used in analyzing the EUI and AIA observations to the modeled expanding loop. When we apply the method to the loop model that
has been convolved with the AIA PSF, however, the measured expansion is reduced to only a factor of 1.7. 

\begin{figure}[ht!]
  \centerline{%
    \includegraphics[width=0.5\textwidth]{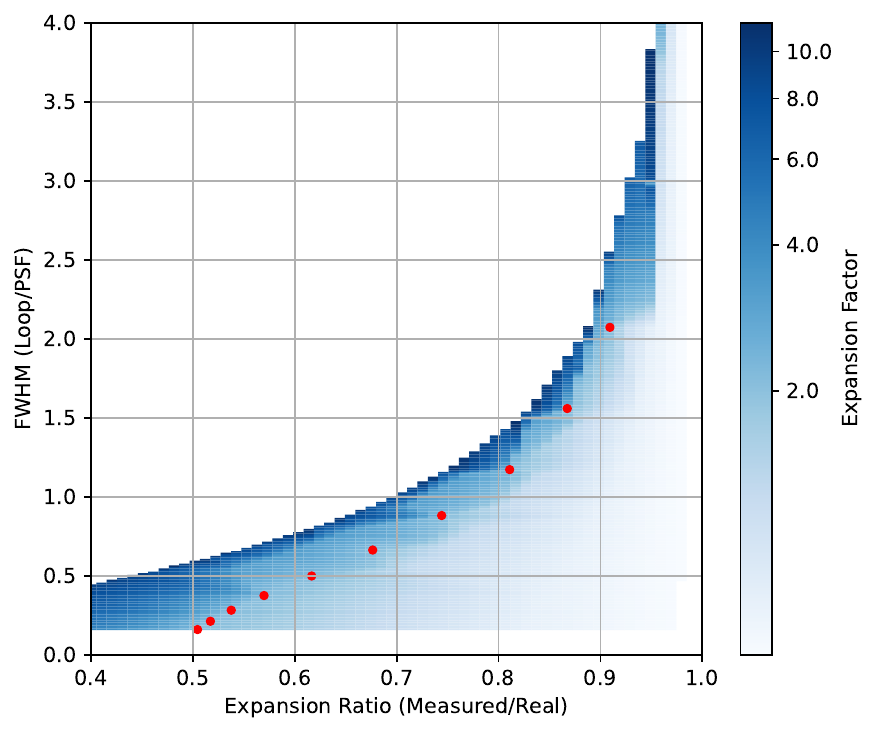}} %
\caption{ Parameter survey comparing the ratio of the loop/PSF FWHM with the measured/actual expansion factor using the model for Figure \ref{fig8}.
The input expansion factor is shown by the color bar. The red dots show the values for the EUI loop in Figure \ref{fig8}.
\label{fig9}}
\end{figure}

Clearly the expansion detected by instruments
with even larger PSFs will be even less. The EUV Imaging Spectrometer \cite[EIS,][]{Culhane2007} on {\it Hinode}, for example, has a PSF of several arcseconds, so the
expansion would likely not be detected at all. To explore the play off between PSF FWHM and loop expansion factor, we performed a wider parameter survey
on our model. The survey consists of taking the loop model from Figure \ref{fig8}, applying a wide range of expansion factors and PSF widths, then measuring the 
expansion using the same technique as before. 
Figure \ref{fig9} shows the results of the survey. We plot the ratio of the loop FWHM to the PSF FWHM against the ratio of the measured to actual (real)
expansion factors. The input expansion factor is shown with the color bar. The red dots show the results for the EUI loop in Figure \ref{fig3} with
an expansion factor of 2.2.

There are a few conclusions that we can draw from the survey. First, 
note that the results skew to the left as the ratio of loop/PSF width
decreases and the actual expansion factor increases. There is a tendency for the data points to accumulate on the left edge of the surface plot
as the expansion factor increases. This implies that there is a limit to how bad the measurement can be if the loop width and expansion is large. This is as
we would expect: if the loop width is much larger than the PSF width then the loop expansion will be well measured.
Second, as the PSF width increases, the measurement becomes less accurate (again note the skew to the left). The effect is more pronounced as the
expansion factor increases (light$\rightarrow$dark color). Third, if we assume that an expansion of 30\% is a reasonable limit to whether it would be detected or not
(assuming that intensity variations on this scale might be comparable to the photometric calibration uncdertainty), then the expansion factor of 2.2 
for the EUI loop in Figure \ref{fig3} that the survey is modeled on would not be detected at all if the PSF width is larger than the loop width. This is comparable
to moving from actual EUI to AIA observations (not idealised models of the PSF). 

\section{Discussion}
Previous work has suggested that coronal loops have nearly constant cross-sections along their full lengths. These observations contradict our expectation 
that they should expand with height as the coronal magnetic field diverges. It is also a widespread view that loops are composed of a number of finer 
unresolved magnetic threads. If expansion occurs below the resolution of the observing instrument, we would not detect it. 

We have argued that a relatively small number of threads is needed to reproduce the widths of coronal loops observed by instruments such as SDO/AIA.
In this picture, the multi-thread model implies the existence of tens of unresolved threads, not thousands. In general,
AIA does not resolve individual coronal loops, but it can do so on rare occasions. A further implication is that an instrument
with better spatial resolution, such as Solar Orbiter/EUI, should resolve more coronal loops, and that any unresolved
loops would be composed of even fewer threads. 

In this paper, we examine whether loop length could be a determining factor in whether we can observe loop expansion.
We have examined cases of long post-flare and trans-equatorial loops with AIA, and the expansion is clearly visible.
We also explore whether expansion is detectable in shorter AR loops at the higher spatial resolution of EUI. 
We find several cases where loop expansion is captured in the observations.

In summary, the long post-flare and trans-equatorial loops show expansion of a factor $>$ 3 when traced over $>$ 60\,Mm\, segments.
The shorter AR loops observed at higher spatial resolution show expansion of a factor $>$ 2 when traced over $>$ 20\,Mm\, segments.

There remain challenges in understanding other properties of active region loops. They are over-dense compared to static
equilibrium theory and persist longer than the expected cooling time \citep{Klimchuk2006,Reale2014}. They also  
have relatively narrow (near isothermal) cross-field temperature distributions \citep{DelZanna2003,Aschwanden2005,Warren2008}. Simulations of impulsive
heating increase the loop apex density, but imply faster cooling. The multi-thread model of impulsively heated coronal loops also suggests
the temperature distribution should be broad.

Reconciling the observed properties of the solar corona will undoubtedly require us to understand how coronal emission is structured. Loop expansion, for example, plays a critical role in determining the density and temperature evolution of a coronal loop \citep{Reep2022}. As already mentioned, \cite{Malanushenko2022} have suggested that what appear to be coronal loops are actually complex, sheet-like structures, and that observed loops may result from line-of-sight effects. The analysis of stereoscopic observations by \cite{Aschwanden2008b,Aschwanden2008a}, however, suggest that loops have consistent properties when viewed and measured from different vantage points. Still, their observations were made at a relatively small separation angle ($\sim$7$^\circ$) and at modest spatial resolution ($\sim$2 pixels, or 2300\,km). The higher spatial resolution and more varied viewing angles from EUI should provide additional insights into this question. A recent study by \cite{Mandal2024} using EUI and AIA observations found consistent loop characteristics despite a 43$^\circ$ spacecraft separation angle, but this analysis was for a single loop.

Another important finding here relates to the quantity of unresolved magnetic threads. We see clear evidence of thread
separation in the EUI data, consistent with the idea that the number of finer threads is relatively small. 
We also find cases where the loop separates into individual threads and at least one of these strands shows expansion similar
to that in the unseparated loop. 
The separated threads that do expand show expansion of a factor of $>1.5$ on length scales of $>$ 20\,Mm.
In one interesting case, the measured widths of the separated threads also broadly agree with
the width of the unseparated loop. Our results support the multi-thread model of coronal loops, and the view that they are composed of 
only a few finer threads. This is encouraging for the ability of future imaging spectrometers such as Solar-C/EUVST \citep{Shimizu2020}
and MUSE \citep{DePOntieu2020} to resolve 
coronal structures and measure their fundamental properties.

\begin{acknowledgments}
The work of D.H.B. and H.P.W. was funded by the NASA Hinode program. 
Solar Orbiter is a mission of international cooperation between ESA and NASA, operated by ESA.
The EUI instrument was built by CSL, IAS, MPS, MSSL/UCL, PMOD/WRC, ROB, LCF/IO with funding from the Belgian Federal Science Policy Office (BELSPO/PRODEX); the Centre National d’Etudes Spatiales (CNES); the UK Space Agency (UKSA); the Bundesministerium für Wirtschaft und Energie (BMWi) through the Deutsches Zentrum für Luft- und Raumfahrt (DLR); and the Swiss Space Office (SSO).
The AIA data are courtesy of NASA/SDO and the AIA, EVE, and HMI science teams.
\end{acknowledgments}

\facilities{SDO, SolO}

\end{document}